\documentclass[aps,twocolumn,preprintnumbers,amsmath,amssymb,floats,citeautoscript,superscriptaddress,nobalancelastpage]{revtex4-1}
\usepackage{graphicx}
\usepackage{dcolumn}
\usepackage{bm}
\usepackage{color}
\newcommand{\Yoshimi}[1]{{\color{black}#1}} 
\newcommand{\Motoyama}[1]{{\color{black}#1}} 
\newcommand{\Hashimoto}[1]{{\color{black}#1}} 
\newcommand{\Kobayashi}[1]{{\color{black}#1}}

\newcommand{\vect}[1]
{{\mbox{\boldmath $#1$}}}

\begin{document}


\title{Dimer-Mott and charge-ordered insulating states in the quasi-one-dimensional
\\organic \Kobayashi{salts} $\delta'_{P}$- and $\delta'_{C}$-(BPDT-TTF)$_2$ICl$_2$}

\author{R.~Kobayashi}
\affiliation{Institute for Materials Research, Tohoku University, Aoba-ku, Sendai 980-8577, Japan}
\author{K.~Hashimoto}
\email[Corresponding author: ]{hashimoto@imr.tohoku.ac.jp}
\affiliation{Institute for Materials Research, Tohoku University, Aoba-ku, Sendai 980-8577, Japan}
\author{N.~Yoneyama}
\affiliation{Graduate Faculty of Interdisciplinary Research, University of Yamanashi, Kohu, Yamanashi 400-8511, Japan}
\author{K.~Yoshimi}
\affiliation{The Institute for Solid State Physics, The University of Tokyo, Kashiwa, Chiba 277-8581, Japan}
\author{Y.~Motoyama}
\affiliation{The Institute for Solid State Physics, The University of Tokyo, Kashiwa, Chiba 277-8581, Japan}
\author{S.~Iguchi}
\affiliation{Institute for Materials Research, Tohoku University, Aoba-ku, Sendai 980-8577, Japan}
\author{Y.~Ikemoto}
\affiliation{Japan Synchrotron Radiation Research Institute, SPring-8, Sayo, Hyogo 679-5198, Japan}
\author{T.~Moriwaki}
\affiliation{Japan Synchrotron Radiation Research Institute, SPring-8, Sayo, Hyogo 679-5198, Japan}
\author{H.~Taniguchi}
\affiliation{Graduate School of Science and Engineering, Saitama University, Saitama 338-8570, Japan}
\author{T.~Sasaki}
\affiliation{Institute for Materials Research, Tohoku University, Aoba-ku, Sendai 980-8577, Japan}


\begin{abstract}
{We investigated the electronic states of the quasi-one-dimensional organic \Kobayashi{salts} $\delta'_{P}$-(BPDT-TTF)$_2$ICl$_2$ and $\delta'_{C}$-(BPDT-TTF)$_2$ICl$_2$, both of which are insulating at room temperature owing to strong electron correlations.
Through measurements of electrical resistivity, optical conductivity, and magnetic susceptibility, as well as band-structure calculations, we have revealed that the two materials possess completely different ground states, even though they have the same chemical composition and stacking configuration of the donor molecules.
We have found that the $\delta_P'$-type salt with an effective half-filled band behaves as a dimer-Mott insulator and exhibits a \Kobayashi{phase transition to a nonmagnetic state} at 25 K, whereas the $\delta'_C$-type salt with a 3/4-filled band shows a charge ordering transition just above room temperature and becomes nonmagnetic below 20 K.
The optical spectra of the $\delta_P'$-type salt are composed of two characteristic bands due to intra- and interdimer charge transfers, supporting the dimer-Mott insulating behavior arising from the strong on-site Coulomb interaction. By contrast, in the $\delta'_C$-type salt, a single band characterizing the formation of charge ordering arising from the off-site Coulomb interactions is observed. Upon lowering the temperature, the shape of the optical spectra in the $\delta_C'$-type salt becomes asymmetric and shifts to much lower frequencies, suggesting the emergence of domain-wall excitations with fractional charges expected in a one-dimensional charge-ordered chain.
The temperature dependence of the magnetic susceptibility of the $\delta_P'$-type salt is well described by a two-dimensional (2D) spin-1/2 Heisenberg antiferromagnetic model on an anisotropic square lattice in the dimerized picture, while in the $\delta_C'$-type salt, it can be explained by a 2D spin-1/2 Heisenberg antiferromagnetic model on an anisotropic honeycomb lattice formed in the charge-ordered state. These completely different ground states between the $\delta'_P$- and $\delta'_C$-type salts come from the difference of degree of dimerization of two face-to-face BPDT-TTF molecules.}
\end{abstract}

\maketitle


\section{Introduction}

Low-dimensional organic \Kobayashi{compounds} exhibit a rich variety of intriguing physical phenomena, such as superconductivity, magnetism, metal--insulator transition, and Dirac fermion behavior, depending on the dimensionality, detailed electronic structure, and electron-correlation strength of the system \cite{Miyagawa04,Dressel04,Kajita14}. Many organic \Kobayashi{compounds} have low-dimensional electronic structures owing to the anisotropic molecular orbitals, which can be widely tuned by the stacking configurations of the constituent molecules. In addition, the characteristic narrow band relative to the Coulomb repulsion, which is formed by overlapping $\pi$ orbitals of the constituent molecules, gives rise to strong electron-correlation effects. Organic \Kobayashi{compounds} therefore have provided an ideal platform for studying strongly correlated low-dimensional systems.

\begin{figure*}[t]
\includegraphics[width=0.95\linewidth]{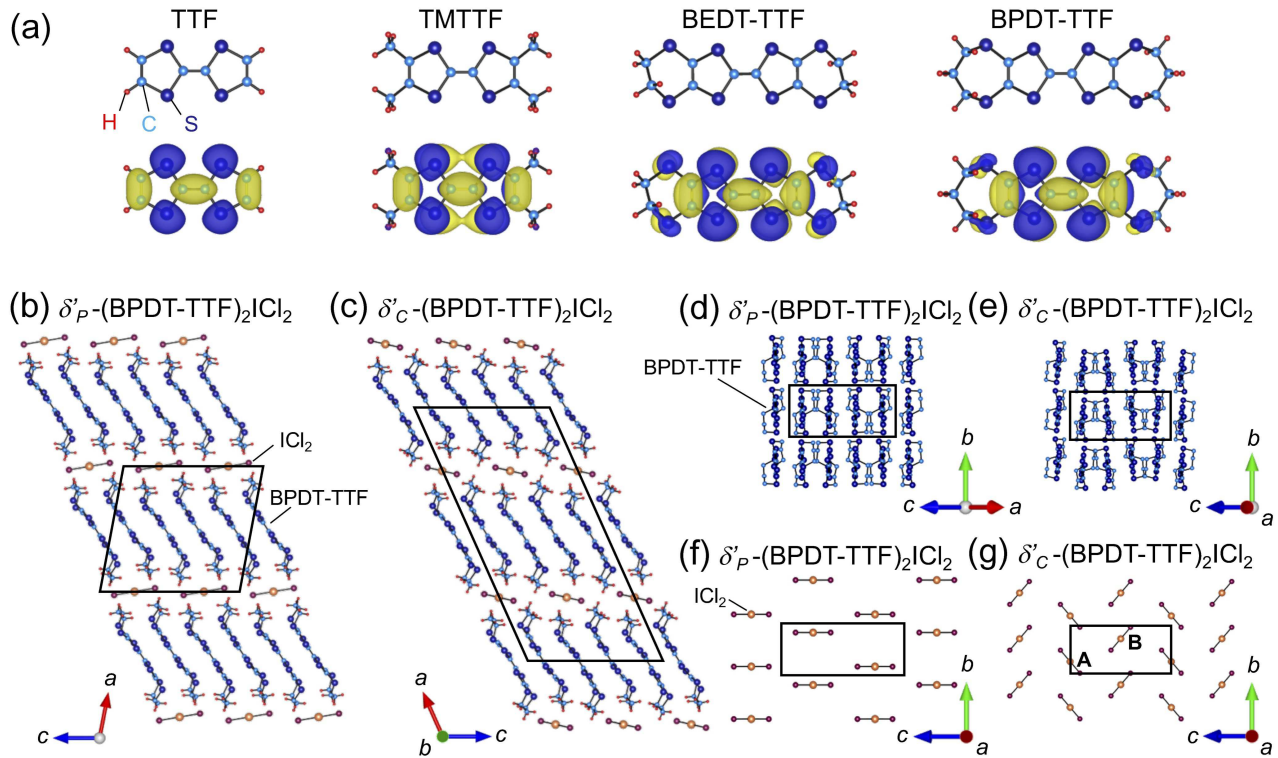}
\caption{(Color online). (a) Representative donor molecules, TTF, and its derivatives. The HOMOs calculated for each molecular unit at the B3LYP/6-31G(d) level \Kobayashi{\cite{gaussian16}} are also shown. (b) and (c) Crystal structures of (b) $\delta^{'}_P$-ICl$_2$ and (c) $\delta^{'}_C$-ICl$_2$. The quadrangles in (b) and (c) represent the unit cell. (d) and (e) Two-dimensional (2D) planes of the BPDT-TTF molecules for (d) $\delta^{'}_P$-ICl$_2$ and (e) $\delta^{'}_C$-ICl$_2$ viewed along the molecular long axis of BPDT-TTF. (f) and (g) Anion configurations of (f) $\delta^{'}_P$-ICl$_2$ and (g) $\delta^{'}_C$-ICl$_2$ viewed along the $a$ direction. The rectangles in (d)--(g) are the projection of the unit cell on the plane perpendicular to the viewing angle.
} \label{optical}
\end{figure*}

The history of organic conductors started with the synthesis of TTF- and TCNQ-based salts with metallic conductivity \cite{Jerome04}, where TTF (tetrathiafulvalene) serves as a donor molecule and TCNQ (tetracyanoquinodimethane) acts as an accepter molecule. Since then, a number of TTF and TCNQ derivatives were designed as donor and acceptor molecules \cite{Geiser04}. Among them, TMTTF (tetramethyltetrathiafulvalene) is a representative donor molecule, which forms quasi-one-dimensional (quasi-1D) charge-transfer salts (TMTTF)$_2$$X$, where $X$ is a monovalent anion \cite{Jerome04}. In this system, face-to-face stacked 1D columnar structures are favored because the highest occupied molecular orbital (HOMO) spreads out mainly perpendicularly to the molecular plane [see Fig.\,1(a)].
In contrast, BEDT-TTF-based charge transfer salts, (BEDT-TTF)$_2$$X$, have quasi-two-dimensional (quasi-2D) electronic structures \cite{Hotta03,Seo04}, where BEDT-TTF is bis(ethylenedithio)tetrathiafulvalene. In this system, columnar structures are usually not preferred because the introduction of the outer six-membered rings including sulfur atoms extends the HOMO distribution parallel to the molecular plane, which leads to a rich variety of 2D staking patterns classified by Greek letters, such as $\alpha$, $\beta$, $\delta$, $\kappa$, $\theta$, etc. \cite{Hotta03,Seo04,Mori98,Mori99-1,Mori99-2}. 
Thus, the wide variety of donor molecules in organic systems has advanced our explorations of novel physical phenomena in strongly correlated low-dimensional systems.

BPDT-TTF [bis(propylenedithio)tetrathiafulvalene] is an analog of BEDT-TTF, in which the terminal ethylene groups of BEDT-TTF are replaced by propylene groups [see Fig.\,1(a)] {\cite{Geiser04,Shibaeva04}. Although the HOMO distribution of the BPDT-TTF molecule is similar to that of the BEDT-TTF molecule [see Fig.\,1(a)], 1D columnar structures are formed in the BPDT-TTF-based charge transfer salts because the steric effects due to the propylene groups prevent 2D molecular arrangements as in the case of the BEDT-TTF-based salts. Syntheses of several BPDT-TTF-based charge-transfer salts have been reported so far \cite{Kobayashi84,Mori84-1,Kato84,Nigrey86,Williams87,Shibaeva88,Yakushi88,Geiser88,Kushch89,Takihirov91}, but the electronic properties have not yet been investigated in detail. 

In this study, we focus on two kinds of BPDT-TTF-based salts with a monovalent ICl$_2$ anion {\cite{Williams87,Shibaeva88}. As shown in Figs.\,1(b) and 1(c), the crystal structures of the two salts consist of alternating stacks of BPDT-TTF and ICl$_2$ layers. The space groups are $P2/c$ [Figs.\,1(b), 1(d), and 1(f)] and $C2/c$ [Figs.\,1(c), 1(e), and 1(g)], respectively \cite{Williams87,Shibaeva88}. Based on the systematic classification of molecular arrangement in the (BEDT-TTF)$_2$$X$ system, the stacking configurations of the donor molecules for both salts are assigned to the $\delta^{'}$ type {\cite{Mori98,Mori99-1,Mori99-2}. We therefore call the former $\delta^{'}_{P}$-(BPDT-TTF)$_2$ICl$_2$ (denoted as $\delta^{'}_{P}$-ICl$_2$) and the latter $\delta^{'}_{C}$-(BPDT-TTF)$_2$ICl$_2$ (denoted as $\delta^{'}_{C}$-ICl$_2$) \cite{note}.
As shown in Figs.\,1(d) and 1(e), the donor planes of both salts consist of 1D columnar stackings of the BPDT-TTF molecules along the $c$ axis. \Hashimoto{In contrast, the anion configurations are different from each other; in $\delta^{'}_{P}$-ICl$_2$ the anion layer consists of the ICl$_2$ molecules parallel to each other [Fig.\,1(f)], while in $\delta^{'}_{C}$-ICl$_2$, ICl$_2$ molecules A and B are related by the 2$_1$ screw axis along the $b$ direction [Fig.\,1(g)].} The charge transfer between the BPDT-TTF layers and the insulating ICl$_2$ layers gives rise to +0.5 hole per one BPDT-TTF molecule, leading to a 3/4-filled band system.
According to the band theory, both salts should be metallic; however, the electrical resistivity shows insulating behaviors \cite{Williams87}. Up to now, the origin of the insulating states has not been investigated. In this study, in order to clarify the detailed nature of the insulating states in $\delta^{'}_{P}$-ICl$_2$ and $\delta^{'}_{C}$-ICl$_2$, we investigated the electronic states of the two salts by measuring electrical resistivity, optical conductivity, and magnetic susceptibility, as well as by performing band-structure calculations. We have found that $\delta^{'}_{P}$-ICl$_2$ with an effective 1/2-filled band behaves as a dimer-Mott insulator and exhibits a \Kobayashi{transition to a nonmagnetic state} at 25 K, whereas $\delta^{'}_{C}$-ICl$_2$ with a 3/4-filled band shows a charge-ordering (CO) transition at around room temperature and becomes nonmagnetic below 20 K. 

The rest of the paper is as follows. We briefly present the experimental details in Sec.\,II. The electronic structures of $\delta^{'}_{P}$-ICl$_2$ and $\delta^{'}_{C}$-ICl$_2$ are discussed based on the band-structure calculations and the electrical resistivity measurements in Sec.\,III A. The results of the optical conductivity for $\delta^{'}_{P}$-ICl$_2$ and $\delta^{'}_{C}$-ICl$_2$ are presented in Secs.\,III B and III C, respectively, and the magnetic properties are presented in Sec.\,III D. Finally, our main results are summarized in Sec.\,IV.

\section{Experiment}
Single crystals of $\delta^{'}_P$-ICl$_2$ and $\delta^{'}_C$-ICl$_2$ were grown through an electrochemical oxidation method {\cite{Williams87,Shibaeva88}.
The crystal shapes of both salts were rectangular. The crystal longitudinal axis of $\delta^{'}_P$-ICl$_2$ corresponds to the stacking direction of the BPDT-TTF molecules, that is, the $c$ axis, while in $\delta^{'}_C$-ICl$_2$, the direction of the minor axis corresponds to the $c$ axis. Although these crystals were mixed in the same batch, we distinguished the two by measuring polarized optical reflectivity for each crystal. The density functional theory (DFT) calculations of the HOMO orbital for each donor molecule at the B3LYP/6-31G(d) level were performed with GAUSSIAN 16 \Kobayashi{\cite{gaussian16}}. Based on the crystal data determined by single-crystal x-ray diffraction at 298 K, we performed band-structure calculations for both salts with the extended H$\ddot{\rm{u}}$ckel method by using the tight-binding approximation \cite{Mori84-2}. The temperature-dependent resistivity measurements were carried out with the standard four-probe method with a current flow along the $c$ axis \Hashimoto{for both salts}, which is the stacking direction of BPDT-TTF molecules [see Figs.\,1(d) and 1(e); \Hashimoto{we cut the crystals of $\delta^{'}_C$-ICl$_2$ perpendicularly to the $b$ axis and made electrical contacts along the $c$ axis]}. The electrical contacts were made using carbon paste. Polarized reflectivity measurements were performed with a Fourier transform microscope spectrometer in the midinfrared region (600--8000 cm$^{-1}$). In the far-infrared region (150--700 cm$^{-1}$), synchrotron radiation light at BL43IR in SPring-8 was used. The reflectivity in the range 8000--45\,000 cm$^{-1}$ was measured at room temperature. The optical conductivity spectra were calculated through a Kramers--Kronig (KK) transformation by assuming the standard high-frequency extrapolation $\omega^{-4}$. The magnetic susceptibility was measured on a superconducting quantum interference device magnetometer (Quantum Design MPMS-XL) \Kobayashi{using several pieces of single crystals. The mass of the crystals used for the magnetic susceptibility measurements of $\delta^{'}_{P}$-ICl$_2$ and $\delta^{'}_{C}$-ICl$_2$  was 2.5 and 0.40 mg, respectively. The magnetic fields were applied along the $b$ axis for both salts.} The Curie-tail contribution at low temperatures and the core diamagnetism were subtracted. 

\section{Results and Discussion}

\subsection{Electronic band structures and insulating states of $\delta^{'}_P$-ICl$_2$ and $\delta^{'}_C$-ICl$_2$}

\begin{figure}[t]
\includegraphics[width=1\linewidth]{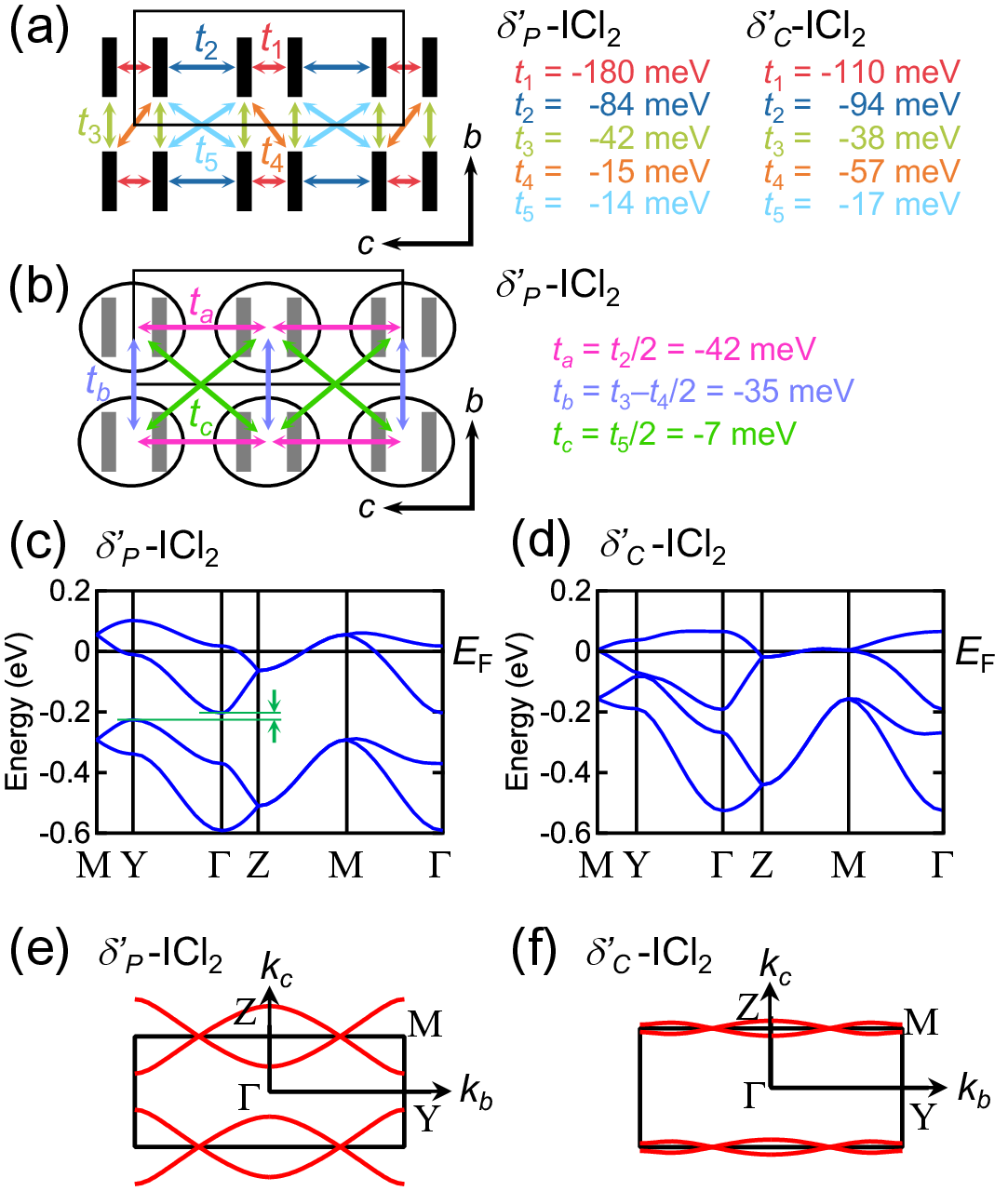}
\caption{(Color online). (a) Dominant transfer integrals $t_1$--$t_5$ in $\delta^{'}_P$-ICl$_2$ and $\delta^{'}_C$-ICl$_2$. The solid rectangles indicate the BPDT-TTF molecules. The black square frame represents the unit cell. (b) Effective transfer integrals $t_a$--$t_c$ in a dimer model for $\delta^{'}_P$-ICl$_2$. The ovals represent dimer units. (c) and (d) Electronic band structures of (c) $\delta^{'}_P$-ICl$_2$ and (d) $\delta^{'}_C$-ICl$_2$ evaluated by tight-binding calculations based on the extended H\"uckel method. (e) and (f) Fermi surfaces of (e) $\delta^{'}_P$-ICl$_2$ and (f) $\delta^{'}_C$-ICl$_2$.
}
\end{figure}

To investigate the electronic structures of $\delta^{'}_P$-ICl$_2$ and $\delta^{'}_C$-ICl$_2$, we performed band-structure calculations with the extended H$\ddot{\rm{u}}$ckel method by using the tight-binding approximation \cite{Mori84-2}. In Fig.\,2(a), the obtained dominant transfer integrals for $\delta^{'}_P$-ICl$_2$ and $\delta^{'}_C$-ICl$_2$ are summarized. Figures\,2(c) and 2(d) show the electronic band structure of $\delta^{'}_P$-ICl$_2$ and $\delta^{'}_C$-ICl$_2$, respectively. As shown in Fig.\,2(d), in $\delta^{'}_C$-ICl$_2$, a 3/4-filled band is formed. By contrast, in $\delta^{'}_P$-ICl$_2$, a clear dimerization gap due to strong dimerization of two face-to-face BPDT-TTF molecules emerges between the antibonding and bonding bands, leading to an effective 1/2-filled band system [see Fig.\,2(c) and also \Hashimoto{the inset of Fig.\,4(b)} below]. The effective model to describe the electronic state at half filling is therefore a so-called dimer model \cite{Koretsune14}, in which each dimer is regarded as a unit with one hole [Fig.\,2(b)]. Figures\,2(e) and 2(f) display the Fermi surfaces of $\delta^{'}_P$-ICl$_2$ and $\delta^{'}_C$-ICl$_2$, respectively. Both salts have a pair of quasi-1D Fermi surfaces perpendicular to the $c$ axis corresponding to the stacking direction of the BPDT-TTF molecules. The band-structure calculations suggest that both compounds have a quasi-1D electronic structure conductive along the stacking direction of the BPDT-TTF molecules \cite{anisotropy of rho}.

\begin{figure}[t]
\includegraphics[width=0.75\linewidth]{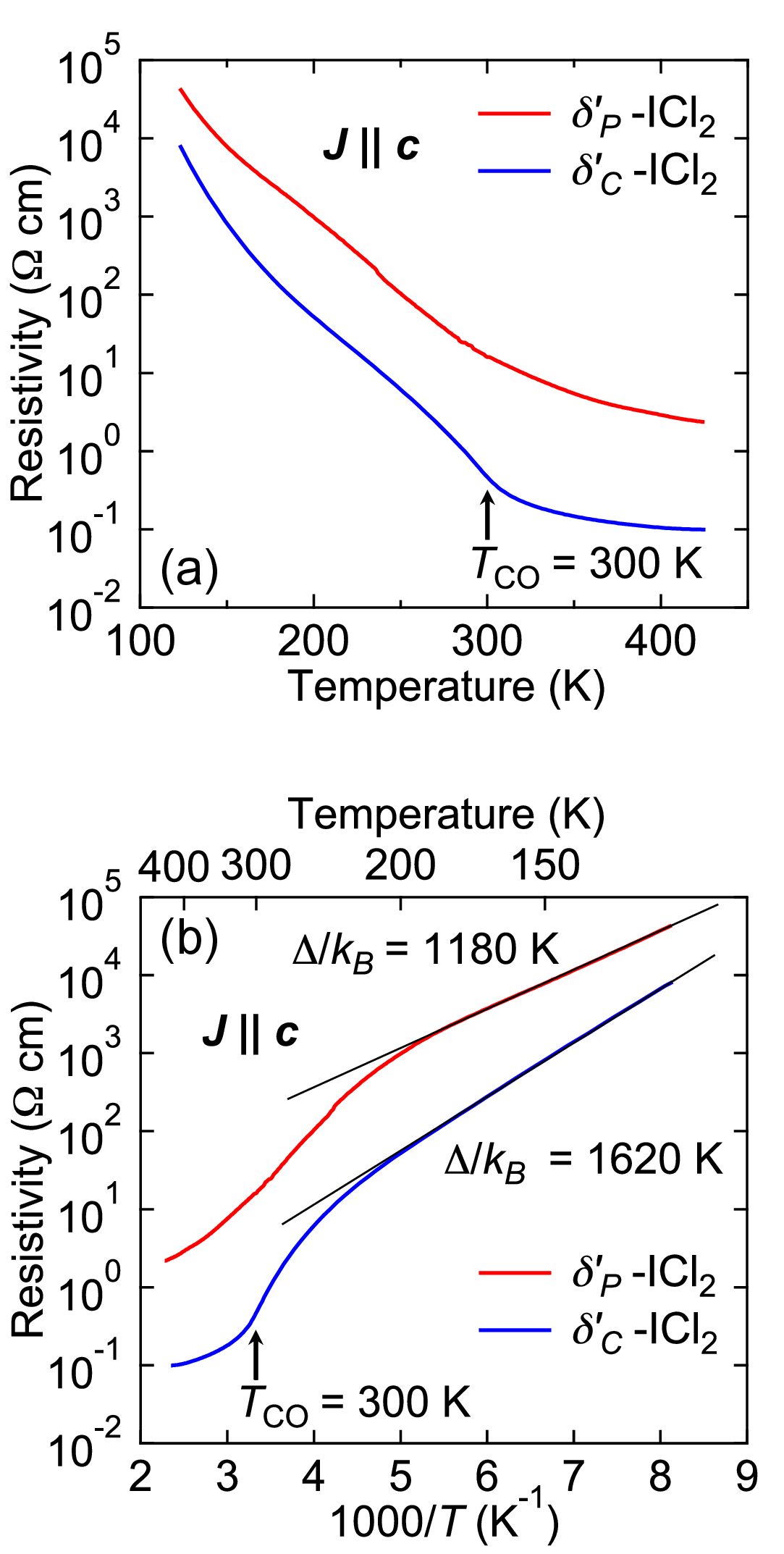}
\label{resistivity}
\caption{(Color online). (a) Temperature \Kobayashi{dependence} of the resistivity of $\delta^{'}_P$-ICl$_2$ and $\delta^{'}_C$-ICl$_2$ \Hashimoto{along the $c$ axis (${\vect{J} \parallel \vect{c}}$)}. (b) Arrhenius plot of the resistivity of $\delta^{'}_P$-ICl$_2$ and $\delta^{'}_C$-ICl$_2$. The solid lines are fits to the Arrhenius law $\propto \exp(\frac{\Delta}{k_B T})$ with $\Delta/k_B = 1180$ and 1620 K for $\delta^{'}_P$-ICl$_2$ and $\delta^{'}_C$-ICl$_2$, respectively.
}
\end{figure}

Figure\,3(a) shows the temperature dependence of the electrical resistivity of $\delta^{'}_P$-ICl$_2$ and $\delta^{'}_C$-ICl$_2$ \Hashimoto{along the $c$ axis}. Although the band-structure calculations suggest that both compounds become metallic, the resistivity shows insulating behaviors. The resistivity of $\delta^{'}_C$-ICl$_2$ exhibits a clear transition at about 300 K. Figure\,3(b) demonstrates the Arrhenius plot of the resistivity of $\delta^{'}_P$-ICl$_2$ and $\delta^{'}_C$-ICl$_2$, showing clear activation behaviors with gaps of 1180 and 1620 K for $\delta^{'}_P$-ICl$_2$ and $\delta^{'}_C$-ICl$_2$, respectively. These results indicate that both compounds are insulating below room temperature.

\subsection{Quasi-1D dimer-Mott insulating state of $\delta^{'}_P$-ICl$_2$}

\begin{figure}[t]
\includegraphics[width=0.85\linewidth]{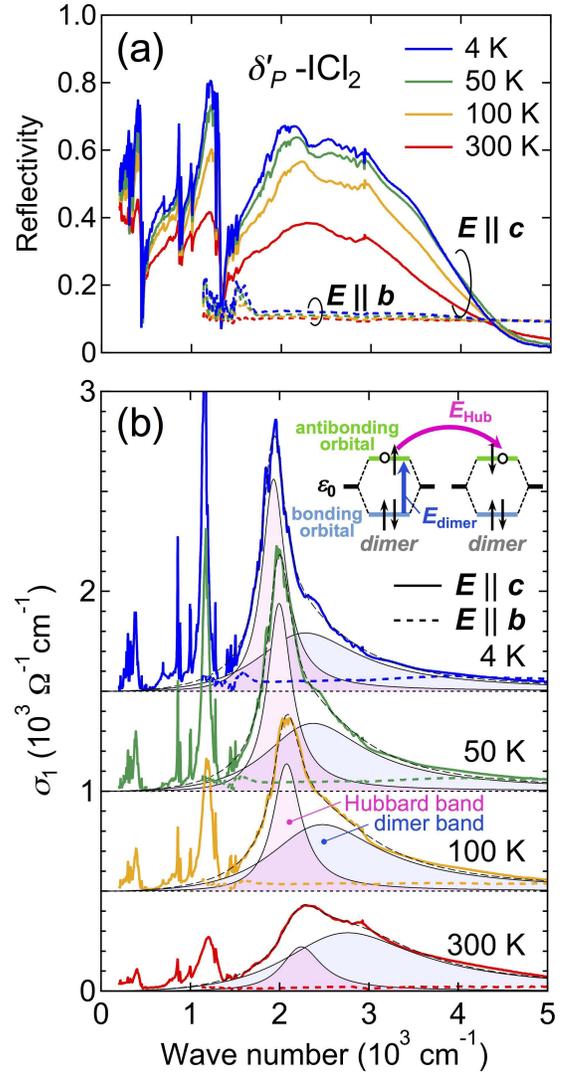}
\label{infrared}
\caption{(Color online). (a) Optical reflectivity and (b) conductivity spectra of $\delta^{'}_P$-ICl$_2$ for ${\vect{E} \parallel \vect{b}}$ and ${\vect{E} \parallel \vect{c}}$ at various temperatures. \Hashimoto{The black solid lines} in (b) represent the Hubbard and dimer bands. \Hashimoto{The black dashed lines show the sum of the Hubbard and dimer bands.} The spectra in (b) are offset for clarity. \Hashimoto{Inset: schematics of intra- and interdimer charge transfer excitations in a dimer-Mott insulator.}
}
\end{figure}

In order to elucidate the nature of the insulating states of $\delta^{'}_P$-ICl$_2$ and $\delta^{'}_C$-ICl$_2$, we performed polarized reflectivity measurements. Figure\,4(a) shows the optical reflectivity spectra $R(\omega)$ of $\delta^{'}_P$-ICl$_2$ for the polarizations of ${\vect{E} \parallel \vect{b}}$ and ${\vect{E} \parallel \vect{c}}$ at various temperatures. A large anisotropy of the reflectivity is clearly observed, reflecting the strong 1D nature of the electronic structure of $\delta^{'}_P$-ICl$_2$. Several vibrational features can be seen below 1500 cm$^{-1}$ for ${\vect{E} \parallel \vect{c}}$, which are attributed to the totally symmetric $a_g$ vibrational modes of the BPDT-TTF molecule.
In the midinfrared region (1500--5000 cm$^{-1}$), the reflectivity spectra show broad features in the polarization of ${\vect{E} \parallel \vect{c}}$. The corresponding optical conductivity spectra $\sigma_1(\omega)$ obtained from the KK transformation are plotted in Fig.\,4(b). At high temperatures, $\sigma_1(\omega)$ for ${\vect{E} \parallel \vect{c}}$ has a broad and asymmetric shape in the range of 1500--5000 cm$^{-1}$. However, upon lowering the temperature, a clear peak structure centered at 2000 cm$^{-1}$ emerges, which makes it clearer that the spectra at 4 K for ${\vect{E} \parallel \vect{c}}$ consist of two bands located at 2000 and 2500 cm$^{-1}$. In addition, a clear optical gap of about 1500 cm$^{-1}$ is observed in the polarizations of ${\vect{E} \parallel \vect{c}}$, which is in good agreement with $2\Delta/k_B$ observed in the dc resistivity [see Fig.\,3(b)]. 

A so-called dimer-Mott insulator, a particular class of Mott insulator with one hole localized per each pair of dimerized molecules, emerges in an effective half-filled band system owing to the strong on-site Coulomb interaction. In a dimer-Mott insulator, there are two absorption bands in the midinfrared region \cite{Faltermeier07,Dumm09,Hashimoto15,Naka13}: the so-called Hubbard and dimer bands corresponding to the inter- and intradimer charge transfers, respectively [see \Hashimoto{the inset of Fig.\,4(b)}]. The excitation energy of the interdimer charge transfer $E_{\rm{Hub}}$ is given by the effective on-dimer Coulomb interaction $U_{\rm{dimer}}$, which causes splitting of the antibonding band.
The excitation energy of the intradimer charge transfer $E_{\rm{dimer}}$ is given by $2 t_d$, where $t_d$ is the intradimer transfer integral. Our band-structure calculations based on the lattice parameters and internal coordinates of $\delta^{'}_P$-ICl$_2$ obtained at room temperature have estimated the intradimer charge transfer $2 t_d = 2 t_1$ to be 360 meV (which corresponds to approximately 2900 cm$^{-1}$), indicating that the broader band located at $\sim$2800 cm$^{-1}$ at room temperature comes from the intradimer charge transfer. The band at 2000--2200 cm$^{-1}$ is therefore attributed to the Hubbard band. It should be noted that the Hubbard band in $\delta^{'}_P$-ICl$_2$ is much sharper than that of other 2D dimer-Mott insulators such as the $\kappa$-type BEDT-TTF salts \cite{Faltermeier07,Dumm09} because of the narrow density of states in the 1D Mott insulating state of $\delta^{'}_P$-ICl$_2$. Very similar optical conductivity spectra have been observed in the 1D dimer-Mott organic insulator $\beta^{'}$-(BEDT-TTF)$_2$ICl$_2$ \cite{Hashimoto15}.

\Hashimoto{Here we note that although $\delta^{'}_P$-ICl$_2$ behaves as a dimer-Mott insulator at low temperatures, the resistivity does not follow the Arrhenius law at high temperatures above 200 K [see Fig.\,3(b)]. This result may suggest the existence of CO fluctuations arising from a possible 3/4 filling nature in $\delta^{'}_P$-ICl$_2$ at high temperatures, as discussed in the next section for $\delta^{'}_C$-ICl$_2$. In order to investigate the presence or absence of CO fluctuations in $\delta^{'}_P$-ICl$_2$, we need further investigation such as synchrotron radiation x-ray diffraction measurements.}

\subsection{Quasi-1D charge-ordered state of $\delta^{'}_C$-ICl$_2$}

\begin{figure}[t]
\includegraphics[width=1.0\linewidth]{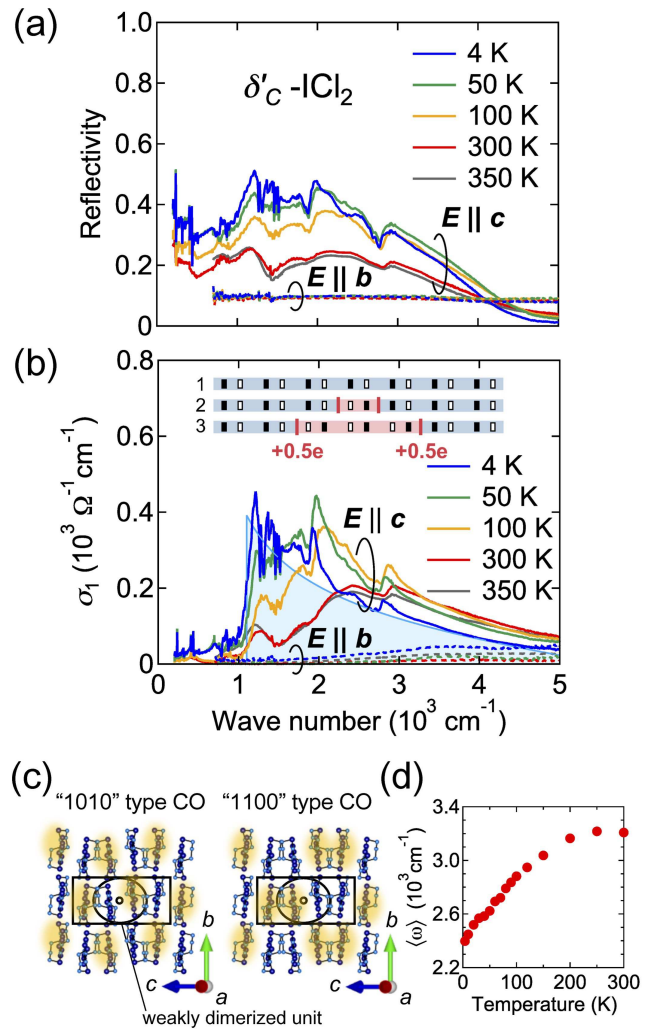}
\caption{(Color online). (a) Optical reflectivity and (b) conductivity spectra of $\delta^{'}_C$-ICl$_2$ for ${\vect{E} \parallel \vect{b}}$ and ${\vect{E} \parallel \vect{c}}$ at various temperatures. The \Kobayashi{light blue} solid curve in (b) represents an asymmetric shape expected for domain-wall excitations. The inset shows a domain-wall excitation with fractional charge $+0.5e$ in a 1D CO chain. (c) Illustrations of the possible CO patterns with a twofold periodicity along the $b$ axis, in which the inversion symmetry of the weakly dimerized unit is broken. The ovals indicate the weakly dimerized units. The small open circles represent the inversion center in the absence of CO. The yellow ovals represent the charge-rich sites. (d) Temperature dependence of the center of gravity of $\sigma_1(\omega)$.
}
\end{figure}

Figures\,5(a) and 5(b) show the polarized optical reflectivity and conductivity spectra of $\delta^{'}_C$-ICl$_2$, respectively. The observed large anisotropy of the optical spectra reflects the strong 1D electronic structure of $\delta^{'}_C$-ICl$_2$. The optical conductivity spectra $\sigma_1(\omega)$ for ${\vect{E} \parallel \vect{c}}$ show several $a_g$ vibrational modes in the region 1200--1400 cm$^{-1}$, which are much broader than those of $\delta^{'}_P$-ICl$_2$. This is due to weaker dimerization in $\delta^{'}_C$-ICl$_2$. In $\delta^{'}_C$-ICl$_2$, in addition to these vibrational modes, several characteristic dip structures are observed in the region 2400--2800 cm$^{-1}$, which can be attributed to the overtones of the fundamental vibrations observed at 1200--1400 cm$^{-1}$. These dip structures are assigned to the C=C stretching $a_g$ modes of the BPDT-TTF molecule coupled with electronic excitations through the electron-molecular vibration interaction \cite{Bozio87}. The presence of the overtones indicates that the energy potential of the molecular vibrations is anharmonic, showing that charge disproportionation accompanied by CO exists between the two weakly dimerized BPDT-TTF molecules \cite{Yamamoto11}. Indeed, recent synchrotron radiation x-ray diffraction measurements have revealed the presence of superlattices corresponding to a double-period lattice modulation along the $b$ axis \cite{KEK}. The possible CO patterns with the twofold periodicity perpendicular to the BPDT-TTF stacking direction, which we call the ``1010"- and ``1100"-type patterns, are shown in Fig.\,5(c). 

\Hashimoto{The 1100-type pattern has been discussed in terms of CO accompanied by bond distortion such as bond-charge-density waves \cite{Seo04,Clay17}. In this case, the spin-singlet formation at the CO transition is expected, which is inconsistent with the experimental result that the magnetic susceptibility of $\delta^{'}_C$-ICl$_2$ does not show any signature of a nonmagnetic ground state at $T_{\rm{CO}}=300$ K, as discussed in the next section.} Therefore, we can exclude the possibility that the 1100-type CO has been realized in $\delta^{'}_C$-ICl$_2$.}

To \Hashimoto{investigate the possibility of realization of the 1010-type CO in} $\delta^{'}_C$-ICl$_2$, we \Hashimoto{examined} a broad band observed in the midinfrared region (1000--5000 cm$^{-1}$) characterizing the CO state. Figure\,5(d) shows the temperature dependence of the center of gravity of $\sigma_1(\omega)$ defined by $\langle \omega \rangle = \int_{1000}^{6000}{\sigma_1(\omega)\omega d\omega}/\int_{1000}^{6000}{\sigma_1(\omega) d\omega}$, which shifts to lower frequencies with decreasing temperature. Generally, in a CO state, a transition between Hubbard-like bands induced by the off-site Coulomb repulsion $V$, i.e., a site-to-site transition within the CO pattern, gives rise to a broad band in the midinfrared region of the order of $V$ \cite{Dressel03,Drichko06,Kaiser10,Hashimoto14}, which is almost temperature independent. In the present system, however, the center of the spectral weight shifts to much lower frequencies with decreasing temperature [see Fig.\,5(d)]. In addition, the shape of the spectra becomes asymmetric with decreasing temperature [see Fig.\,5(b)]. In the 1010-type CO pattern, domain-wall excitations with fractional charge $+0.5e$ are theoretically predicted as elementary excitations [see the inset of Fig.\,5(b)] \cite{Mayr06,Fratini07}. The energy of the fractional charge excitations is characterized by not $V$ but $V/t$, where $t$ is the transfer integral. In organic \Kobayashi{compounds}, $t$ increases as the temperature decreases owing to the thermal contraction of the lattice. This effect may cause the large reduction in $V/t$ in $\delta^{'}_C$-ICl$_2$. The large energy shift of the optical spectra in $\delta^{'}_C$-ICl$_2$ therefore implies the existence of domain-wall excitations with fractional charges. It should be noted that very similar behavior has been observed in quasi-1D charge-ordered organic \Kobayashi{salts} $\delta$-(EDT-TTF-CONMe$_2$)$_2$$X$ \cite{Antal13}.

\Hashimoto{Here we note that the fact that the overtones are observed even at 350 K in $\delta^{'}_C$-ICl$_2$ [Figs.\,5(a) and 5(b)] suggests the presence of CO fluctuations above $T_{\rm{CO}}$ as observed in $\alpha^{'}$-(BEDT-TTF)$_2$IBr$_2$, where a possible short-range CO has been discussed in terms of a charge-order-disorder phase transition not accompanied by structural changes \cite{Yue09}.}

\subsection{Magnetic properties of $\delta^{'}_P$-ICl$_2$ and $\delta^{'}_C$-ICl$_2$}

\begin{figure}[t]
\includegraphics[width=1.0\linewidth]{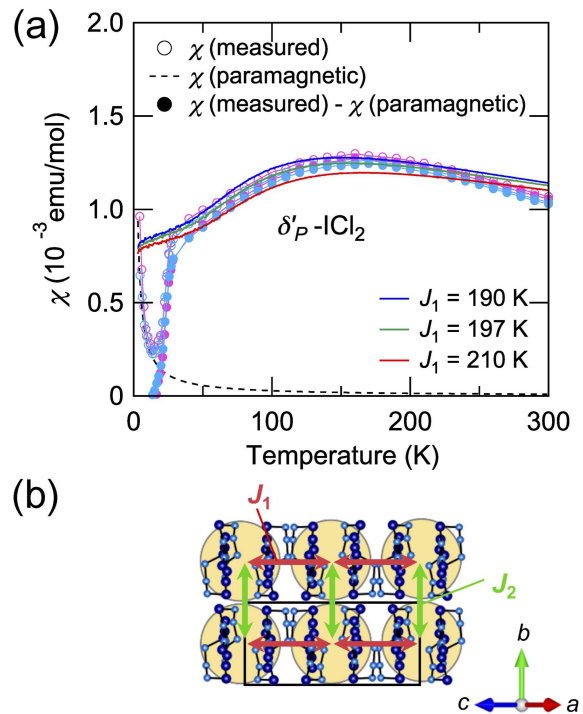}
\caption{(Color online). (a) Temperature dependence of the magnetic susceptibility of $\delta^{'}_P$-ICl$_2$ \Hashimoto{measured at magnetic fields of 0.5 T (magenta circles) and 2 T (light blue circles) parallel to the $b$ axis. The open circles represent the experimental data \Hashimoto{after subtracting the core diamagnetism}. The dotted line shows a Curie component due to a small amount of magnetic impurities, $\sim$ 1.0 \%. The solid circles are the data obtained by subtracting the contribution of the magnetic impurities. The solid curves show the results of the QMC calculations based on \Yoshimi{a 2D spin-1/2 Heisenberg antiferromagnetic model on an anisotropic square lattice with  $J_1=190, 197, 210$ K and $J_2/J_1=0.675$.}} (b) Dominant antiferromagnetic interactions $J_1$ and $J_2$ in the dimer model for $\delta^{'}_P$-ICl$_2$.}
\end{figure}

In order to investigate the magnetic properties of $\delta^{'}_{P}$-ICl$_2$ and $\delta^{'}_{C}$-ICl$_2$, we performed magnetic susceptibility measurements. 
\Yoshimi{We also calculated the magnetic susceptibility by using the ALPS/LOOPER code \cite{alps2007,alps2011,alpsweb,alps-looper,alps-looper-web}. By performing quantum Monte Carlo (QMC) calculations with several lattice sizes up to 1024 spins for $\delta^{'}_{P}$-ICl$_2$, we confirmed that the finite-size effect does not remain.} Figure\,6(a) shows the temperature dependence of the magnetic susceptibility $\chi(T)$ of $\delta^{'}_{P}$-ICl$_2$ \Kobayashi{for $\vect{H} \parallel \vect{b}$}, which shows a broad hump structure at high temperatures, followed by a sharp drop at around 25 K. 
We evaluated \Kobayashi{$\chi(T)$ of $\delta^{'}_{P}$-ICl$_2$} by considering the dimer approximation, as shown in Fig.\,6(b).
Because the effective transfer integrals $t_{a}$ and $t_{b}$ are given by $t_2/2$ and $t_3-t_4/2$, respectively, \Yoshimi{$J_{1}=4t_{a}^2/U_{\rm{dimer}}$ and $J_{2}=4t_{b}^2/U_{\rm{dimer}}$} are estimated to be \Yoshimi{197 and 133 K, where $U_{\rm{dimer}} = 276$ meV} was extracted from the optical conductivity measured at room temperature [see Sec.\,III B]. 
\Kobayashi{Above 25 K, $\chi(T)$ is well described by a 2D spin-1/2 Heisenberg antiferromagnetic model on an anisotropic square lattice with $J_1=197$ K and $J_2=133$ K calculated by the QMC method. }
The sharp drop at 25 K \Hashimoto{has been observed also in $\chi(T)$ for $\vect{H} \parallel \vect{a^{\ast}}$ ($\perp bc$) and $\vect{H} \parallel \vect{c}$ (neither of which are is shown),} showing the emergence of a nonmagnetic ground state below 25 K. Such a nonmagnetic ground state can be attributed to the formation of spin singlets between the two neighboring spins along the $c$ axis, indicative of the presence of a lattice instability in the 1D system.

\begin{figure}[t]
\includegraphics[width=1.0\linewidth]{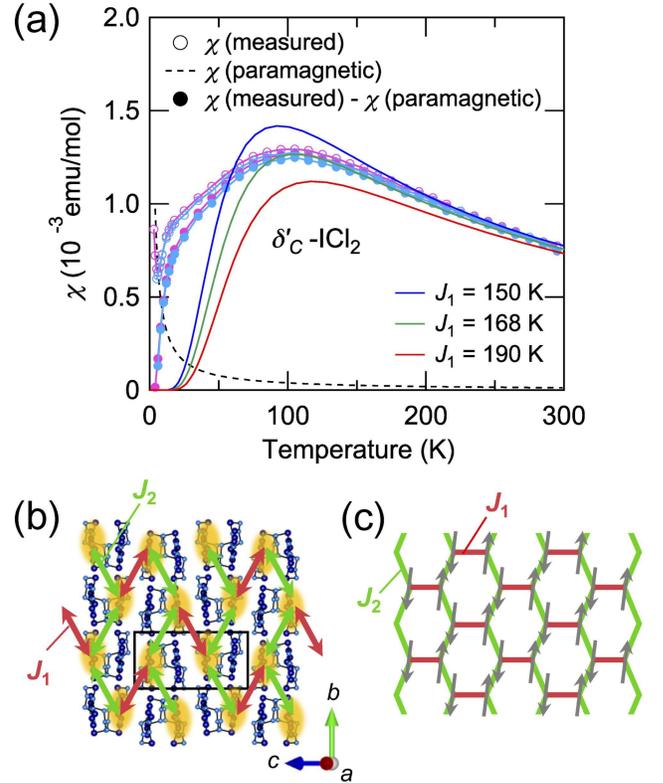}
\caption{(Color online). (a) Temperature dependence of the magnetic susceptibility of $\delta^{'}_C$-ICl$_2$ \Hashimoto{measured at magnetic fields of 1 T (magenta circles) and 5 T (light blue circles) parallel to the $b$ axis. The open circles represent the experimental data \Hashimoto{after subtracting the core diamagnetism}. The dotted line shows a Curie component due to a small amount of magnetic impurities, $\sim$ 0.8 \%. The solid circles are the data obtained by subtracting the contribution of the magnetic impurities. The solid curves show the results of the QMC calculations based on a} spin-1/2 Heisenberg model on an anisotropic honeycomb lattice with \Yoshimi{$J_1=150, 168, 190$ K and $J_2/J_1=0.1$}. (b) Dominant \Kobayashi{antiferromagnetic interactions $J_1$ and $J_2$} in the CO state of $\delta^{'}_C$-ICl$_2$. (c) Illustration of the anisotropic honeycomb lattice in  $\delta^{'}_C$-ICl$_2$. The gray arrows indicate the localized spins with $S=1/2$.
}
\end{figure}

Figure\,\Kobayashi{7(a)} shows $\chi(T)$ of $\delta^{'}_{C}$-ICl$_2$. The position of a broad hump structure shifts to a lower-temperature region compared to that of $\delta^{'}_{P}$-ICl$_2$, suggesting that the formation of CO in $\delta^{'}_{C}$-ICl$_2$ affects the magnetic interactions of the system. Figure\,\Kobayashi{7(b)} represents the dominant antiferromagnetic interactions in the CO state of $\delta^{'}_{C}$-ICl$_2$, which form an anisotropic honeycomb lattice, as shown in Fig.\,\Kobayashi{7(c)}. Because \Yoshimi{$J_1=4 t_4^2/U_0$ and $J_2= 4 t_5^2/U_0$} considering the \Yoshimi{second-order} perturbation processes, $J_1 \gg J_2$ holds in the case of $\delta^{'}_{C}$-ICl$_2$. We calculated the temperature dependence of the magnetic susceptibility expected in a 2D spin-1/2 Heisenberg model on the anisotropic honeycomb lattice \Yoshimi{by the QMC method} \Hashimoto{with the ALPS/LOOPER code \cite{alps2007,alps2011,alpsweb,alps-looper,alps-looper-web}} at \Yoshimi{$J_1 = 150,~168$, and $190$ K, with $J_2/J_1 =  t_5^2/t_4^2 \sim 0.1$}. We confirmed by performing the calculations with lattice sizes up to 2048 spins that the finite-size effect does not remain. As shown in Fig.\,\Kobayashi{7(a)}, the calculations well reproduce the experimental results. In an anisotropic honeycomb lattice, the formation of spin singlets at the strong $J_1$ bonds is expected at a low temperature. Therefore, a rapid drop of $\chi(T)$ in $\delta^{'}_{C}$-ICl$_2$ at about 20 K, which has been observed also in $\chi(T)$ for $\vect{H} \parallel \vect{a^{\ast}}$ and $\vect{H} \parallel \vect{c}$ (neither of which are is shown), can be attributed to the spin-singlet formation between the two neighboring spins along the $J_1$ direction, which is different from that of $\delta^{'}_{P}$-ICl$_2$.

\section{Conclusion}
To summarize, we have performed electrical resistivity, optical conductivity, and magnetic susceptibility measurements as well as band-structure calculations for the quasi-1D organic \Kobayashi{salts} $\delta'_{P}$-ICl$_2$ and $\delta'_{C}$-ICl$_2$. Although the band-structure calculations predict that both compounds should be metallic, strongly correlated insulating states are realized in both compounds owing to Coulomb repulsions. We find that $\delta^{'}_P$-ICl$_2$ behaves as a dimer-Mott insulator and exhibits a \Kobayashi{phase transition to a nonmagnetic state} at 25 K, whereas $\delta^{'}_C$-ICl$_2$ shows a CO transition at around room temperature and becomes nonmagnetic below 20 K. The optical spectra in the dimer-Mott insulator $\delta^{'}_P$-ICl$_2$ are composed of two characteristic bands due to intra- and interdimer charge transfers. By contrast, in $\delta^{'}_C$-ICl$_2$, a single band characterizing CO is observed. Upon lowering the temperature, its shape becomes anisotropic and shifts to much lower frequencies, suggesting the presence of domain-wall excitations with fractional charges expected in a 1D CO chain. The temperature dependence of the magnetic susceptibility of $\delta^{'}_P$-ICl$_2$ is well described by a 2D spin-1/2 Heisenberg antiferromagnetic model on an anisotropic square lattice, while in the $\delta_C'$-type salt, it can be explained by a 2D spin-1/2 Heisenberg model on an anisotropic honeycomb lattice. Even though the two materials have the same chemical composition and stacking configuration of the BPDT-TTF molecules, they possess completely different ground states because of the difference of degree of dimerization. The present system may serve as a model system for studying 1D strongly correlated electron systems.

\section*{Acknowledgments}
We thank M. Naka, A. Ueda, and K. Itoh for fruitful discussions. Synchrotron radiation measurements were performed at SPring-8 with the approval of JASRI (2014B1340, 2015B1756, 2016A0073, and 2016B0073). \Motoyama{The QMC calculation in this work was performed using the facilities of the Supercomputer Center, the Institute for Solid State Physics, the University of Tokyo.} This work was supported by a Grant-in-Aid for Scientific Research (Grants No.\,25287080, No.\,15H00984,  No.\,16K05747, and  No.\,17H05138) from MEXT and JSPS and by a Grant-in-Aid for Scientific Research on Innovative Areas ``$\pi$-Figuration'' (Grant No.\,26102001). \Yoshimi{K.Y. and Y.M. were supported by Building of Consortia for the Development of Human Resources in Science and Technology, MEXT, Japan.}



\begin{thebibliography}{99}
\bibitem{note}
We note that in previous reports, $\delta^{'}_{P}$-ICl$_2$ and $\delta^{'}_{C}$-ICl$_2$ were called $\alpha^{'}$-(BPDT-TTF)$_2$ICl$_2$ and $\alpha$-(BPDT-TTF)$_2$ICl$_2$, respectively, although the stacking configurations of the BPDT-TTF molecules are neither $\alpha^{'}$ type nor $\alpha$ type.
\bibitem{anisotropy of rho}
The resistivity anisotropy at room temperature is $\rho_{a^{\ast}}:\rho_b:\rho_c = 3800:30:1$ for $\delta^{'}_{P}$-ICl$_2$ and $1900 :45 :1$ for $\delta^{'}_{C}$-ICl$_2$, where the $a^{\ast}$ direction is perpendicular to the $bc$ plane.
\end{thebibliography}
\end{document}